\UseRawInputEncoding
\documentclass{aa}  
\usepackage{natbib}
\usepackage{orcidlink}
\usepackage{graphicx}
\usepackage{txfonts}
\begin{document} 

    \title{\centering{\textit{AstroSat} observations  of  interacting  galaxies  NGC~7469 and IC~5283}}

   \author{Abhinna Sundar Samantaray
          \inst{1,2} \thanks{abhinna.samantaray@uni-heidelberg.de}\orcidlink{0000-0002-0635-2264}
           \and H.K. Jassal \inst{2} \thanks{hkjassal@iisermohali.ac.in}\orcidlink{0000-0003-2486-5634}
          \and
          Kulinder Pal Singh \inst{2}\fnmsep\thanks{kps@iisermohali.ac.in}{\orcidlink{0000-0001-6952-3887}}\and G.C. Dewangan \inst{3} \orcidlink{0000-0003-1589-2075}
          }

   \institute{Astronomisches Rechen-Institut (ARI), Universit\"{a}t Heidelberg, M\"{o}nchhofstraße 12-14, Heidelberg 69120, Germany\\
              \email{abhinna.samantaray@uni-heidelberg.de}
         \and
             Department of Physical Sciences, Indian Institute of Science Education and Research Mohali, Knowledge City, Sector 81, SAS Nagar, Punjab 140306, India
         \and Inter-University Centre for Astronomy and Astrophysics, Ganeshkhind, Pune 411007, India 
    }
   \date{Received 18 April}

 
\abstract
{}
{We carry out deep near-ultraviolet (NUV) and far-ultraviolet (FUV) imaging of an interacting galaxy system, comprised of a Seyfert type 1 galaxy NGC~7469 and its companion IC~5283. Our aim is to resolve and map the star-forming regions in the outer arms and look for signs of interaction between the two galaxies.}
{We used {\it AstroSat} Ultra Violet Imaging Telescope (UVIT) to obtain NUV and FUV images of NGC~7469 in a range of filters. We have carried out photometry of star-forming regions in the two galaxies and found their spatial distributions.  We also obtained the distributions of star formation rates (SFR) in NGC~7469 and IC~5283 using the estimates obtained from the FUV and NUV bands. We also carried out Kolmogorov-Smirnov tests to look for differences in the SFRs in the two galaxies. We derived the spectral energy distribution (SED), leading to the determination of physical parameters, including the overall SFR, stellar mass ($\text{M}_{*}$), dust mass ($\text{M}_\text{Dust}$), and specific star formation rates (sSFRs) in both the galaxies.}
{Our NUV and FUV images show the presence of an outer spiral arm that is better resolved. We have identified 33 new star-forming regions out of 51 total identified in the UVIT composite image. Enhanced star formation activity is observed to coincide with the interaction, and KS tests show that there are no significant differences in the SFR distributions of NGC~7469 and IC~5283, indicating that the interaction between the galaxies has not influenced their star formation processes differently. The SED plots and the photometric results demonstrate that most of the star formation activity is confined inside the central starburst (SB) ring.}
{}

   \maketitle
%

\section{Introduction}
Studies of interacting galaxies have made a significant contribution to our understanding of galaxy evolution and the processes that shape their morphology, star formation, and nuclear activity. The gravitational interaction between galaxies can disrupt the equilibrium of the systems due to tidal forces, leading to the emergence of intricate morphologies that deviate from the regular, symmetric shapes observed in isolated galaxies. For example, it can lead to the formation of tidal tails, bridges, and shells \citep{1972ApJ...178..623T, 1982ApJ...252..455S, 1992ARA&A..30..705B}. 
These morphological features provide evidence of the dynamical interaction between galaxies and are often used as indicators of past or ongoing interactions.

The increased gravitational perturbations, gas compression, and shocks associated with interactions can also lead to enhanced star formation activity leading to the formation of starburst regions, where star formation occurs at a rate significantly higher than that observed in isolated galaxies \citep{1996ARA&A..34..749S, 1998ARA&A..36..189K, 2000ApJ...530..660B, 2008ASPC..390..178M}.
Furthermore, interactions between galaxies can influence nuclear activity by driving gas inflows towards the central regions, triggering accretion onto the supermassive black holes that reside at the cores of many galaxies (\citet{1989Natur.340..687H}; \citet{2006ApJS..163....1H}). 
This process can result in the activation of active galactic nuclei (AGN), accompanied by the release of significant amounts of energy across a wide range of wavelengths \citep{2013ARA&A..51..511K}. 
Therefore, by examining these interactions across different wavelengths, we can gain a more comprehensive understanding of the underlying physical processes at play.

One such interacting system consists of a Seyfert 1 galaxy NGC~7469 and its neighbor, IC~5283.  This system is also known as Arp~298. 
The two galaxies have been observed by a number of telescopes at many different wavelengths, over the years and their salient features are described below. 
NGC~7469 is an intermediate spiral galaxy $-$ SABa type.  It is located at a redshift of z = 0.0163 and is one of the brightest (V=13 mag) nearby Active Galactic Nuclei (AGN) and has been studied extensively in multiple wavebands \citep{1997ApJS..113...69W,1998ApJ...505..594N, 2003A&A...403..473K}. 
Optical observations suggest the presence of a bar-like structure (connecting them to the central source) which influences the formation and morphology of the two major spiral outer arms of the galaxy NGC~7469 \citep{2004ARA&A..42..603K, 2004ApJ...602..148D, 2006AJ....131..701L}. The HST has provided high-resolution images that reveal intricate details of these spiral arms, bar, and nucleus \citep{2020MNRAS.493.3656C, 2020ApJ...898...75I}. Dark dust lanes, a consequence of interstellar dust absorbing and scattering light \citet{2001PASP..113.1449C}, are seen in the optical images as well. Nebulae and HII regions within the arms, visible in optical wavelengths, contribute to the overall richness of NGC~7469's appearance.

Earlier observations have also shown the presence of an inner starburst (SB) ring around the nucleus of NGC~7469 which was discovered in radio by \cite{1981ApJ...247..419U} and \cite{ 2003AJ....126..143S} and later imaged in optical and in the infrared  \citep{2007ApJ...661..149D} surrounding an inner circumnuclear disk (CND)  \citep{2004ApJ...602..148D, 2015ApJ...811...39I}. Later observations by ALMA resolved the sizes of the AGN, CND, and SB ring to be about  $1^{''}$, $3^{''}$, and $5^{''}$ from the center \citep{2015ApJ...811...39I, 2020ApJ...898...75I}. These radio observations from ALMA by \citet{Izumi_2020}, accompanied by VLBA observations by \citet{2003ApJ...592..804L} showed the presence of a compact, flat-spectrum radio source in the nucleus of the galaxy, which is a characteristic of Seyfert galaxies and is believed to be associated with accretion onto a central supermassive black hole.  It was discovered as an X-ray source by \citet{1978ApJS...38..357F} based on observations with the UHURU satellite. Further X-ray observations have revealed the presence of a luminous X-ray source in the nucleus, which is possibly related to the accretion process \citep{1989Natur.340..687H, 2006ApJ...643..641H}.

The active nucleus and the inner starburst (SB) region emit predominantly in the IR regime in NGC~7469 \citet{2000AJ....119..991S}. Observations by Galaxy Evolution Explorer (GALEX)\footnote{\url{https://archive.stsci.edu/prepds/galex\_atlas/}} in  FUV and NUV \citet{2007ApJS..173..185G} have reported that the NGC~7469 is one of the Ultra-Violet Luminous Galaxies (UVLGs) according to the criteria mentioned in \citet{2005ApJ...619L..35H}. These observations show very few knot-like features at the northwest side of the galaxy NGC~7469 but with very poor resolution \citep{2007ApJS..173..185G, 2009yCat..21730185G}. Observations with the XMM-Optical Monitor (OM) also show the presence of inner spiral arms ($\approx$ 30 arcsecs), and some fainter outer arms of NGC~7469 ( $\approx$ 60 arcsecs) are also visible, but these two components are unresolvable \citep{2003A&A...403..481B}. In the case of IC~5283, the UV emission is more or less homogeneous, and three distinct parts were seen by the  XMM-OM \citep{2003A&A...403..481B}. 

NGC~7469 exhibits prominent signs of interaction, including tidal features and asymmetrical structures \citep{1995ApJ...438..604M}. 
It seemingly interacts with a neighboring Scd galaxy, IC~5283 which is at the same optical redshift. 
IC~5283 appears to have a knotty distribution and a semi-detached tidal tail \citep{1994AJ....108...90M}.
Together these two galaxies form an isolated pair of interacting galaxies located at $~$66.5 Mpc ($H_0 = 67.8 \text{ km} \text{ s}^{-1} \text{ Mpc}^{-1}$, $\Omega_{m} = 0.308$, $\Omega_{vac}=0.692$) in the constellation of Pegasus known as Arp 298, KPG 575 or Holm 803. Some general properties of these two interacting galaxies aka. NGC~7469 and IC~5283, are summarised in Table \ref{tab:table1}. 
The companion galaxy IC~5283  is located at a projected distance of approximately 50 kpc from NGC~7469  \citep{1995ApJ...438..604M}. 
IC~5283 is classified as a barred spiral galaxy and exhibits distinctive morphological features that provide evidence of its interaction with NGC~7469. 
The presence of tidal features, such as tidal tails or bridges, and asymmetric structures indicate the dynamical influence exerted by the gravitational interaction between the two galaxies \citep{1995ApJ...438..604M}. 
These morphological signatures suggest that IC~5283 has been subjected to significant perturbations, leading to its present appearance.
It has been suggested in \cite{2004ApJS..153...93G} that NGC~7469 is drawing gas from the southeast side of IC~5283 and that the two galaxies are counter-rotating.
Additionally, other studies have explored the impact of the interaction on IC~5283's star formation activity and nuclear properties, shedding light on the interplay between the interacting galaxies \citep{2011A&A...535A..93P, 2017A&A...604A...2S}.

In this paper, we present the UV imaging observations of the pair of galaxies, NGC~7469 and IC~5283, carried out with the Ultra-Violet Imaging Telescope (UVIT) \citep{2017AJ....154..128T} aboard {\it AstroSat} \citep{2014SPIE.9144E..1SS}. 
The higher spatial resolution of the UVIT (1.2$-$1.4 arcsec),  as compared to GALEX (5 arcsecs) helps us gain a more detailed perspective on the star-forming regions within NGC~7469 and its companion galaxy, IC~5283, and discern the intricate structures and properties of these regions.

The paper is organised as follows.  We present the observational details in \S2, followed by the data reduction techniques used and the images obtained in the various near ultraviolet (NUV) and far ultraviolet (FUV) bands in \S3. In \S4, we present further analysis by performing aperture photometry on UV images and identifying star-forming regions of these galaxies. Here we also present the derived parameters of the star-forming regions and the results of performing Kolmogorov-Smirnov (KS) tests on the parameters of the two galaxies to see differences, if any, between the star-forming regions of the two galaxies.  We present the multi-wavelength photometric analysis of the UV data in \S5. 
The spectral energy distributions (SED) of the two galaxies are presented here along with the extracted physical parameters, such as star formation rate (SFR), stellar mass ($\text{M}_{*}$), dust mass ($\text{M}_\text{Dust}$), and specific star formation rate (sSFR) for both galaxies.
In section \ref{discussions}, we summarise our results. The results pertaining to the nuclear activity in NGC~7469 obtained from these observations have already been presented by \citet{2023MNRAS.521.4109K} and \citet{2023ApJ...950...90K}.

\begin{table*}[!h]
	\centering
	\caption{General characteristics of the interacting galaxies NGC~7469 and IC~5283.}
	{
	\begin{tabular}{lcccc} 
 	\hline
	 & NGC~7469 & & IC~5283 & \\
	\hline	\hline
	Properties & Values & References & Values & References\\
	\hline	
	Right Ascension  & $23^{\text{h}}03^{\text{m}}15^{\text{s}}
.6$  & NED & $23^{\text{h}}03^{\text{m}}17^{\text{s}}.6$  & NED \\

	Declination & $+ 08^{\text{d}}52^{\text{m}}26^{\text{s}}$  & NED  & $+ 08^{\text{d}}53^{\text{m}}23^{\text{s}}$  & NED  \\

	Redshift & 0.016317 $\pm$ 0.00001&  NED  & 0.016024 $\pm$  0.000027&  NED \\
 
	Luminosity Distance (Mpc) & 70.2 & NED  & 69.6 & NED  \\	

       Scale (kpc/") & 0.315 & \citet{2000MNRAS.311..120D}  & 0.327 & \citet{2011PhDT.........8V} \\	

   Morphological Type  & (R’)SAB(rs)a  &  NED  & SA(r)cd  & NED \\ 

    Nuclear Activity Type  & Seyfert-1  &  NED,\citet{2008ApJS..174..282L} & N.A. & - \\

   Log ($\text{L}_{\text{dust}}/\text{L}_{\odot}$) &  [11.65]  &  \citet{2023ApJS..265...37Y}  & [11.65] &  \citet{2011PhDT.........8V}\\ 

   log(SFR)  & 1.51&  \citet{2023ApJS..265...37Y}  & 0.50 & \citet{2023ApJS..265...37Y}\\ 

   i $^{\circ}$  & 30.2  & \citet{2020MNRAS.493.3656C}  & 60 & \citet{1994AJ....108...90M} \\ 

   $\text{PA}_{\text{phot}}$($^{\circ}$) & 126  &  \citet{2020MNRAS.493.3656C}  & 120 & \citet{1994AJ....108...90M} \\ 

Stellar age (CND) [Myr] &  110–190 & \citet{2007ApJ...671.1388D}  & 807 & \citet{2011PhDT.........8V} \\
	\hline  
 
	\end{tabular}}
 \label{tab:table1}

   	\tablefoot{ Most of the parameters here are listed in the NASA Extragalactic Data Base, NED operated by the Jet Propulsion Laboratory, California Institute of  Technology, under contract with the National Aeronautics and Space Administration (NASA). The inclination position angle and `i' and `$\text{PA}_{\text{phot}}$', are defined as the inclination between the line of sight and polar axis of the galaxy and the major axis position angle (northwestwards), respectively. `N.A.' means the data is insufficient or not available. `[]' means that the data has been calculated for the Arp 298 i.e. for the interacting system as a whole.}
\end{table*}

\begin{table*}[h]
	\centering
	\caption{Log of Observations of the Seyfert Galaxy NGC~7469 and IC~5283 in the UVIT Filters. These are {N245M, N279N in the NUV, and F154W, F172M in the FUV bands} where the numbers in the middle 245, 279, 154, and 172 correspond to the central wavelength of the corresponding filter in nm. More information on filters is available at: \url{https://uvit.iiap.res.in/Instrument/Filters}}
	\resizebox{2\columnwidth}{!}{
	\begin{tabular}{cccccc} 
		\hline
	Observation ID & UVIT Filter &  Start Time (UT) & Stop Time (UT) & Exposure (s)\\
                     &  & Y:M:D:H:M & Y:M:D:H:M & & \\
		\hline	
  G05\_238T01\_9000001364 &  NUV/N279N &  2017:07:06:19:41 & 2017:07:08:12:16 & 20679 \\ 

  G05\_238T01\_9000001364 & FUV/F154W &  2017:07:06:19:41 & 2017:07:08:12:23 & 22247\\
	
  G08\_071T02\_9000001620 & NUV/N245M & 2017:10:15:20:39 & 2017:10:18:18:31 & 53031\\	

  G08\_071T02\_9000001620 & FUV/F172M & 2017:10:15:20:39 & 2017:10:18:11:59 & 45119 \\
  \hline
	\end{tabular}}
	\label{table2}

\end{table*}

\begin{figure*}
\centering
	\includegraphics[width=0.71\textwidth]{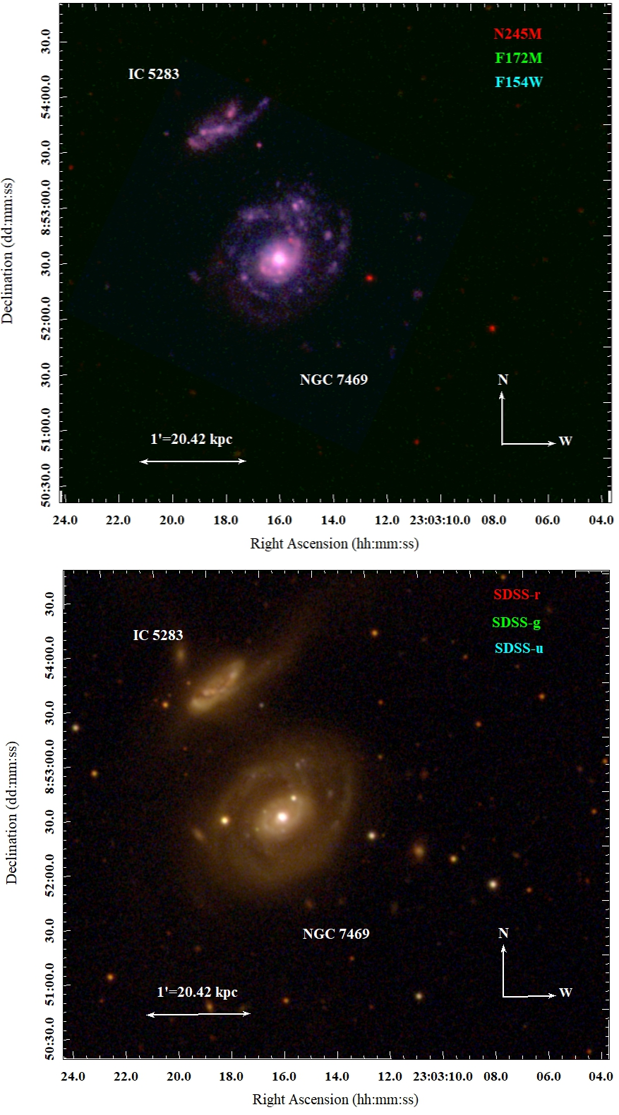}
    \caption{A composite RGB image of NGC~7469 created using the {N245M (red), F172M (green), and F154W (blue)} respectively {\textit(top)}. This image clearly shows the presence of inner spiral arms, and knots which act as evidence of the star formation occurring in the arms of the galaxy.  A composite RGB image of NGC~7469 was created using the SDSS frames,  SDSS-r(red), SDSS-g (green), and SDSS-u (blue) respectively \textit{(bottom)}. Clearly, the spiral arms are better resolved in the UVIT image compared to the SDSS image.}
    \label{Merged_image}
\end{figure*}

\section{Observations }
\label{obs}

Observations of NGC~7469 and IC~5283 were carried out by $AstroSat$ on 7 August 2017 and 15 October 2017 with the UVIT, in the NUV and FUV bands in the Photon Counting mode (see \citep{2017AJ....154..128T, 2020AJ....159..158T} for more details). 
Two $37.5cm$ Ritchey-Chrétien telescopes make up the UVIT, one of which is used to view the far ultraviolet (FUV; range: $130$nm to $180$ nm). 
 The second telescope consists of a dichroic beam splitter that separates the beam into near ultraviolet (NUV: $190$ $-$ $304$ nm) and visible (VIS: $304$ nm $-$ $550$ nm) channels. 
 Each channel consists of many filters as given in \citet{2020AJ....159..158T}. 
 The data from the VIS channel is used to track the spacecraft and is not used for scientific purposes. 

The observations were carried out when the Sun avoidance angle was $\ge 45^{\circ}$ and the RAM angle ($> 12^{\circ}$) where the RAM angle is the angle between the payload axis to the velocity vector direction of the spacecraft. 
UVIT observations and the resulting datasets that were used here are listed in Table \ref{table2}, we list the observations for all the instruments from which data were used in this article, along with the filters that were used for the FUV and NUV observations. The filters used, the start and stop times, and the useful exposure times obtained are also provided.

\section{Data Reduction and Imaging}

We downloaded  Level-1 data from the {\it AstroSat} archive \url{https://astrobrowse.issdc.gov.in/astro_archive/archive/Home.jsp}. 
 The data contained individual orbit-wise FUV and NUV data files as FITS binary tables organized by file type and orbit numbers. The photon counting mode data contain lists of photon detection events with their centroid positions.
The data were analysed using  CCDLAB \citep{2021JApA...42...30P, 2017PASP..129k5002P}. The image reduction included collating all orbit-wise data into individual directories, performing fidelity checks, instrument corrections like field distortions and flat fielding, exposure array corrections, correcting for drifts using the VIS data, and finally applying the World Coordinate System (WCS).\\
The final science products were extracted after co-aligning all the orbit-wise data for all available \textit{AstroSat} observations, thus producing final images in each filter.\\

\section{Analysis and Results}
\label{ana_res}
\subsection{UV images}
A composite image made from three filters, viz., N245M, F172M, and F154W, focusing on the two galaxies is shown in the upper panel of Figure \ref{Merged_image}. The image shows the central bulge, inner arms, and outer spiral arms of NGC~7469.
An optical composite image obtained from the three SDSS filters (r,g, and v) is also shown in the lower panel of this figure. Our observations clearly resolve the two major spiral arms and show the star-forming regions. The image also shows diffuse emission in the outer disc of NGC~7469. 

In the case of the companion galaxy IC~5283, we detect diffuse emission and many star-forming regions via a photometric procedure as described below.

\begin{figure*}
\centering
	\includegraphics[width=1.0\textwidth]{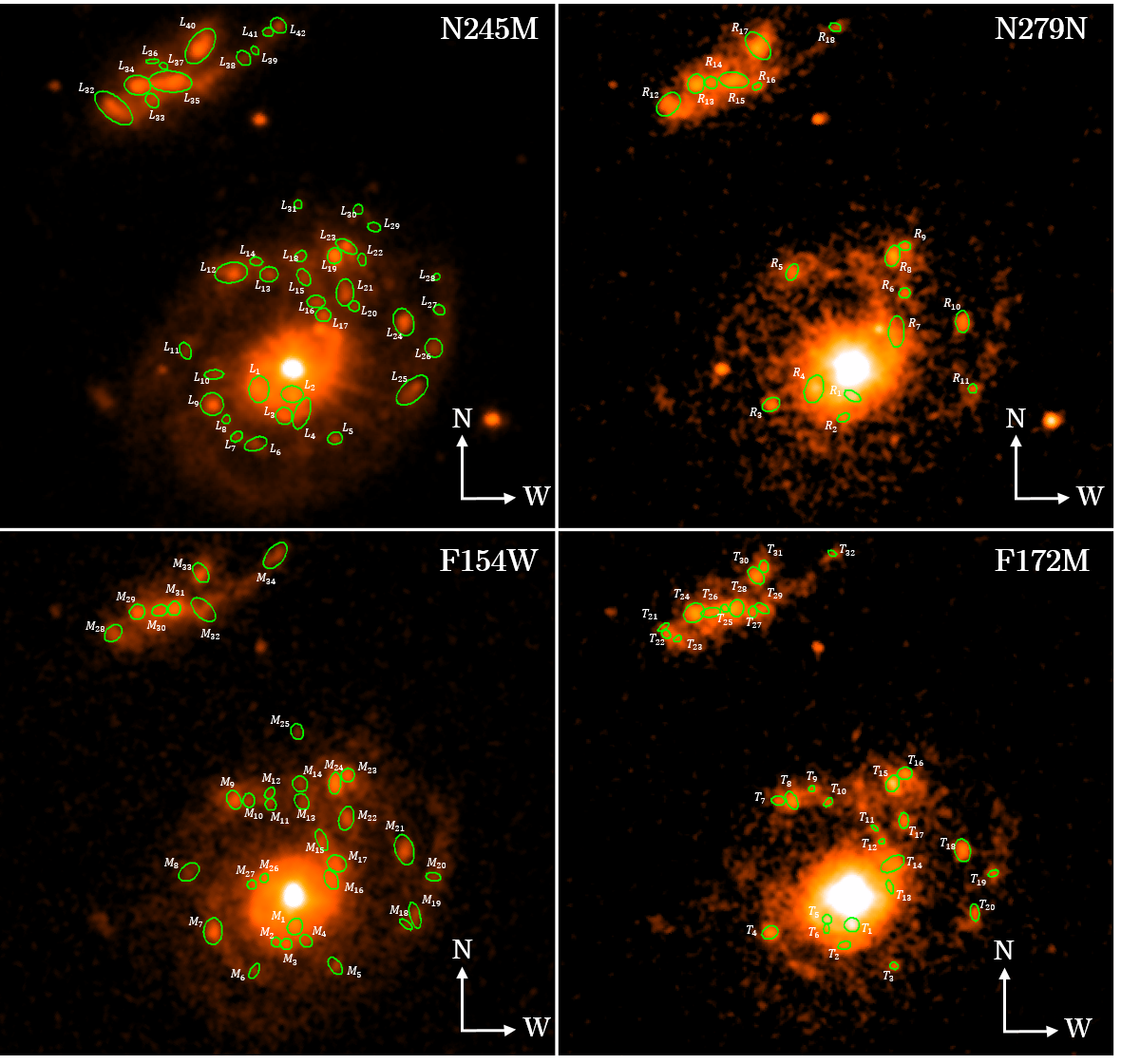}
    \caption{{NGC~7469 and IC 5283 photometry showing the star-forming regions (in green ellipses) in the N245M, F154W, N279N, and F172M filters identified with the help of SExtractor using the parameters mentioned in Table \ref{SExtractorTable}. A similar SF region's identification process was done for all the other UVIT filters as mentioned in the section \ref{NGC7469 photometry}. }}
    \label{Merged_photometry}
\end{figure*}

\begin{table*}
\centering
	\caption{SEXTRACTOR (version v1.2b14) parameters used for the analysis of the UVIT images.}
	{
	\begin{tabular}{lc} 
		\hline
	Main Parameters & Values \\
		\hline	
            \hline
	DETECT\_MINAREA & 10   \\
	DETECT\_THRESH ($\sigma$) & 4 \\	
   FILTER\_NAME  & gauss\_4.0\_7x7.conv   \\ 
   DEBLEND\_NTHRESH  & 32.0  \\ 
   DEBLEND\_MINCONT  & 0.005 \\ 
   CLEAN\_PARAM & 1.0 \\ 
  PHOT\_APERTURES  & 5  \\ 
PHOT\_AUTOPARAMS &  1.5, 1.5  \\
PHOT\_PETROPARAMS &  2.0, 1.5 \\
PIXEL\_SCALE ($''/\text{pix}$) &  0.417 \\
STARNNW\_NAME &  default.nnw \\
BACK\_TYPE &  AUTO \\

	\hline  
\end{tabular}}
\label{SExtractorTable}

\end{table*}

\subsection{UV Photometry}
    \label{NGC7469 photometry}
We performed aperture photometry of the star-forming regions in the galaxies NGC~7469 and IC~5283 using the SEXTRACTOR software developed by \citet{1996A&AS..117..393B}. We will use the `SF regions' for the star-forming regions interchangeably from here on in the entire paper. A detailed procedure for the identification of SF regions, the distribution of their sizes, and the extraction of fluxes of these regions for both galaxies is outlined below.

\subsubsection{Identification of Star-Forming regions}

We extracted the SF regions from a subsection of the image of the size of $427\times 370$ pixels (1 pixel= $0.417$ arcsec) centered on ($\alpha$, $\delta$) = ($23^h 03^m 13.8^s$, $8^{\circ}53^m35.0^s$). The minimum number of pixels for identification of an SF region was fixed to 10 ($\text{DETECT\_MINAREA = 10}$), which is equivalent to the area of a circle of diameter $\sim 3.5$ pixels (approximating the full-width half maximum of the point spread function of the UVIT). We used elliptical shapes for the SF regions.  We  de-blended the SF regions with $32$ subthreshold ($\text{DEBLEND\_THRESHOLD} = 32$) and a minimum contrast parameter value of $0.005$ ($\text{DEBLEND\_MINCONT} = 0.005$). UVIT has a resolution of $\sim 1.4$ arcsec in the FUV and $\sim 1.2$  arcsec in the NUV band which means that UVIT can resolve SF regions in the NGC~7469 down to approximately $0.16 \text{ kpc}^2$ in the FUV and $0.14 \text{ kpc}^2$ in the NUV bands. The background level was automatically detected and SExtractor gives the background subtracted photometric counts of the SF regions (parameter $\text{BACK\_TYPE} \ = \ \text{AUTO}$). We required a minimum flux of $4 \ \sigma$ above the background in a pixel, where $\sigma$ is the average sky noise for a source to be registered. 

These parameters were used to detect the SF regions in all the filters which resulted in the detection of 45, 34, 23, and 32 SF regions in N245M, F154W, N279N, and F172M filters, respectively. The SF regions selected (in the individual filters) are such that there is no overlap between them, thus avoiding any contamination from the adjacent regions. The main parameters used are listed in Table \ref{SExtractorTable} and the spatial distribution of the SF regions detected in various UVIT filters is shown in Figure \ref{Merged_photometry}. The regions $\text{L}_2$, $\text{R}_1$, $\text{M}_1$, and $\text{T}_1$, in the N245M, N279N, F154W, and F172M filter respectively were identified using manual inspection. The reason that these regions were not detected using the SExtractor is mostly due to their high background counts due to diffuse emission and/or due to confusion by unresolved regions. Some regions in close vicinity of the smaller SF regions were not detected by SExtractor in some filters,  for example, $\text{L}_{40}$ is in close vicinity of smaller SF regions ($\text{T}_{30}$ and $\text{T}_{31}$) and were not detected in the N245M filter. However, the radial profile of L40 showed a smaller variation due to the smaller regions. Therefore, the L40 region was corrected by manually encompassing the undetected regions using radial profiles. The regions  $\text{R}_{7}$, $\text{L}_{32}$, $\text{M}_{34}$, and $\text{L}_{35}$ were also corrected for the same reason. All the SF regions' parameters like RA, Dec, parameters for the elliptical region, flux, and SFR, are being made available electronically.

\begin{figure*}
\centering
	\includegraphics[width=1.0\textwidth]{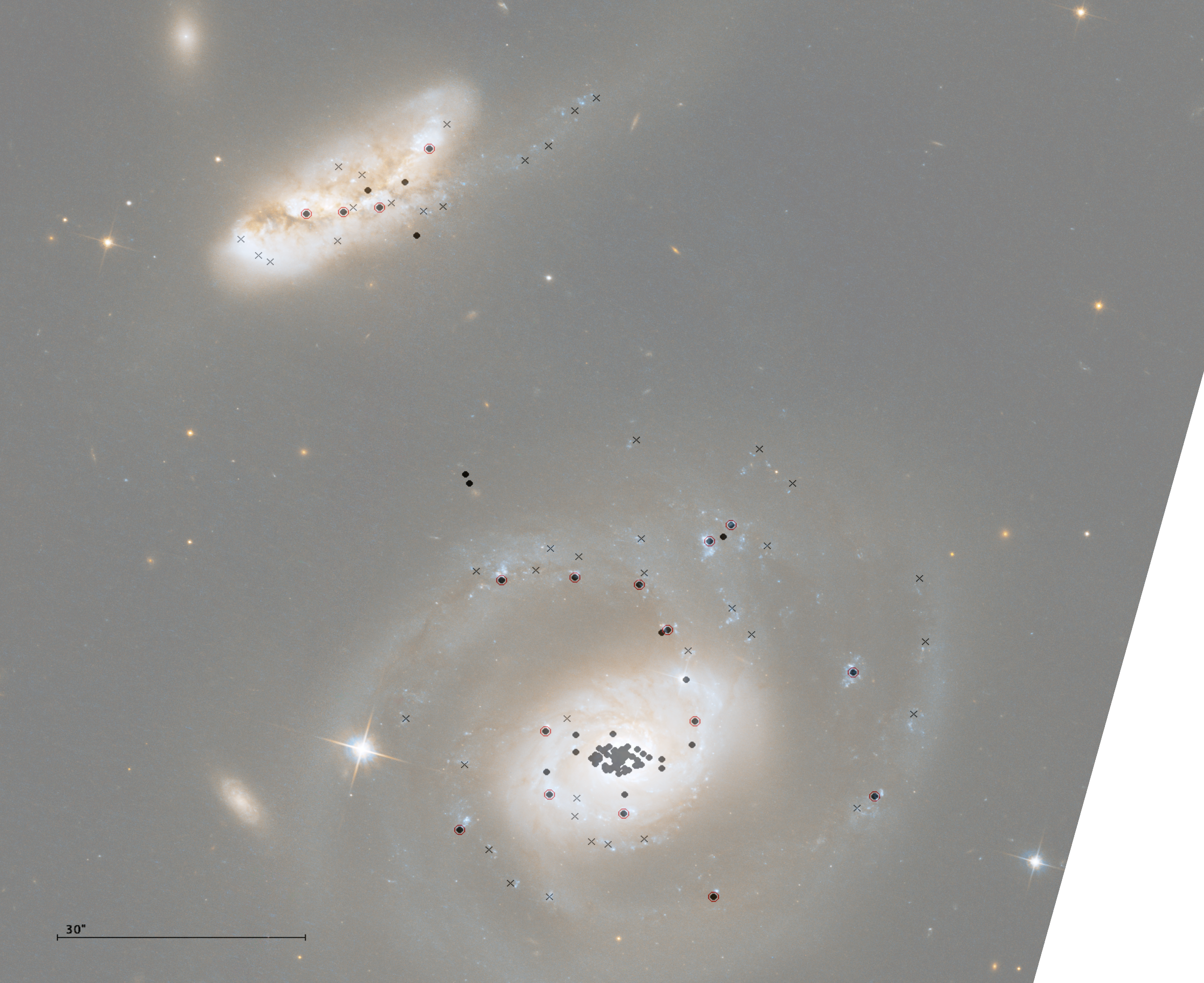}
    \caption{A comparison between new and previously detected clumps plotted on the HST image taken from (\url{https://esahubble.org/}). The previous clump/SF region detections are marked in black dots. We detected 51 unique SF regions identified using RA and Dec in all the UVIT filters. We identified 33 new SF regions which are shown in black crosses whereas the red open circles (enclosing black dots) represent the clumps/SF regions (18) which are detected both by the UVIT and were observed previously. }
    \label{detections}
\end{figure*}

We identified a total of  51  SF regions, out of which 18 regions were detected by the previous observations for both galaxies. We have shown these regions plotted over the HST image in Figure \ref{detections}. Overall, 33 new SF regions were identified as compared to previous observations \citep{2007ApJ...661..149D, 2009MNRAS.399.1641P, 2011ApJ...731...92R}.  

\begin{figure*}
\centering
	\includegraphics[width=1.0\textwidth]{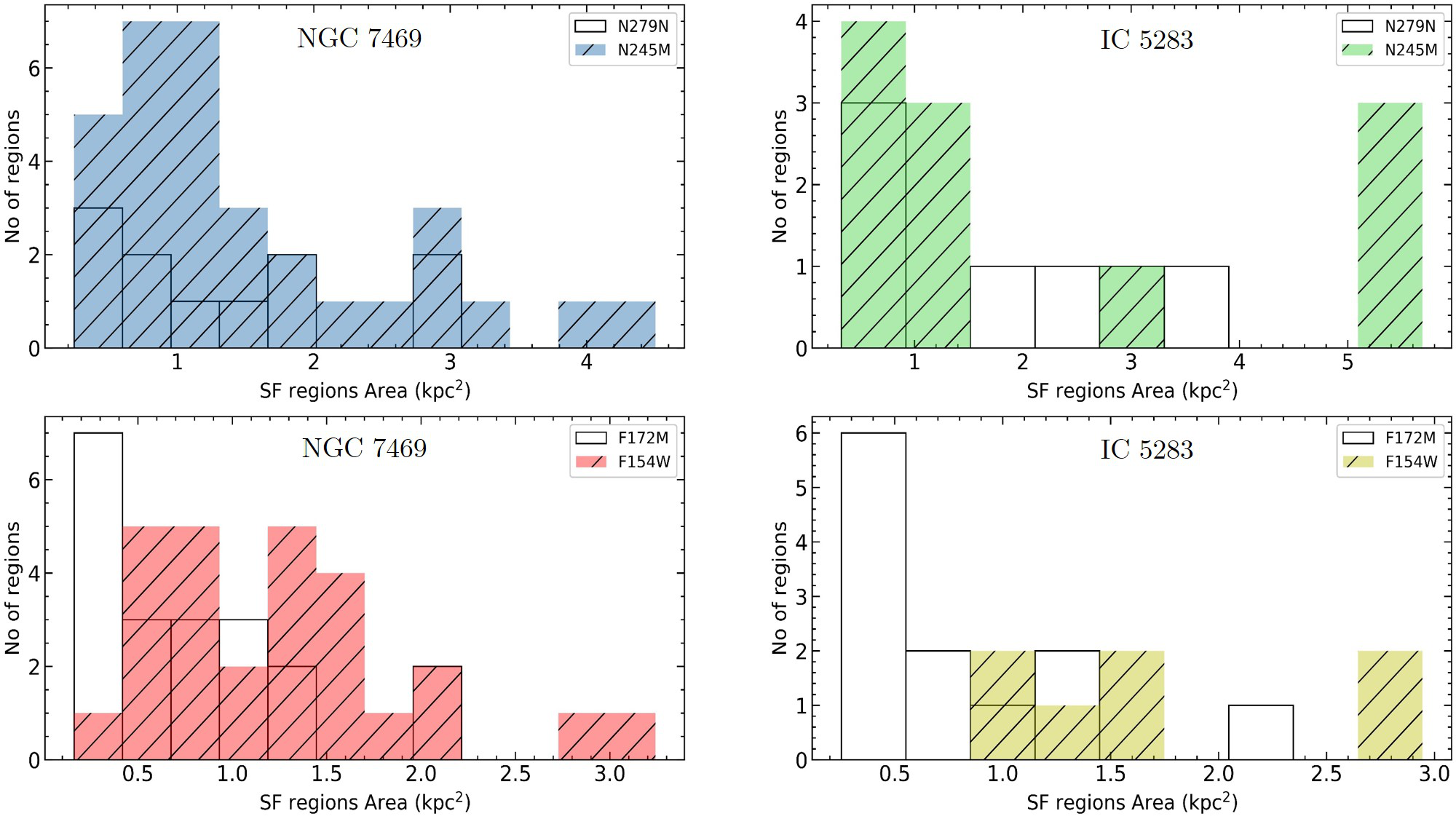}
    \caption{Distribution of areas of the SF regions detected by SExtractor in NGC~7469 and IC~5283 in various UVIT filters}
    \label{Merged_Knots}
\end{figure*}

\begin{figure*}
\centering
	\includegraphics[width=1.0\textwidth]{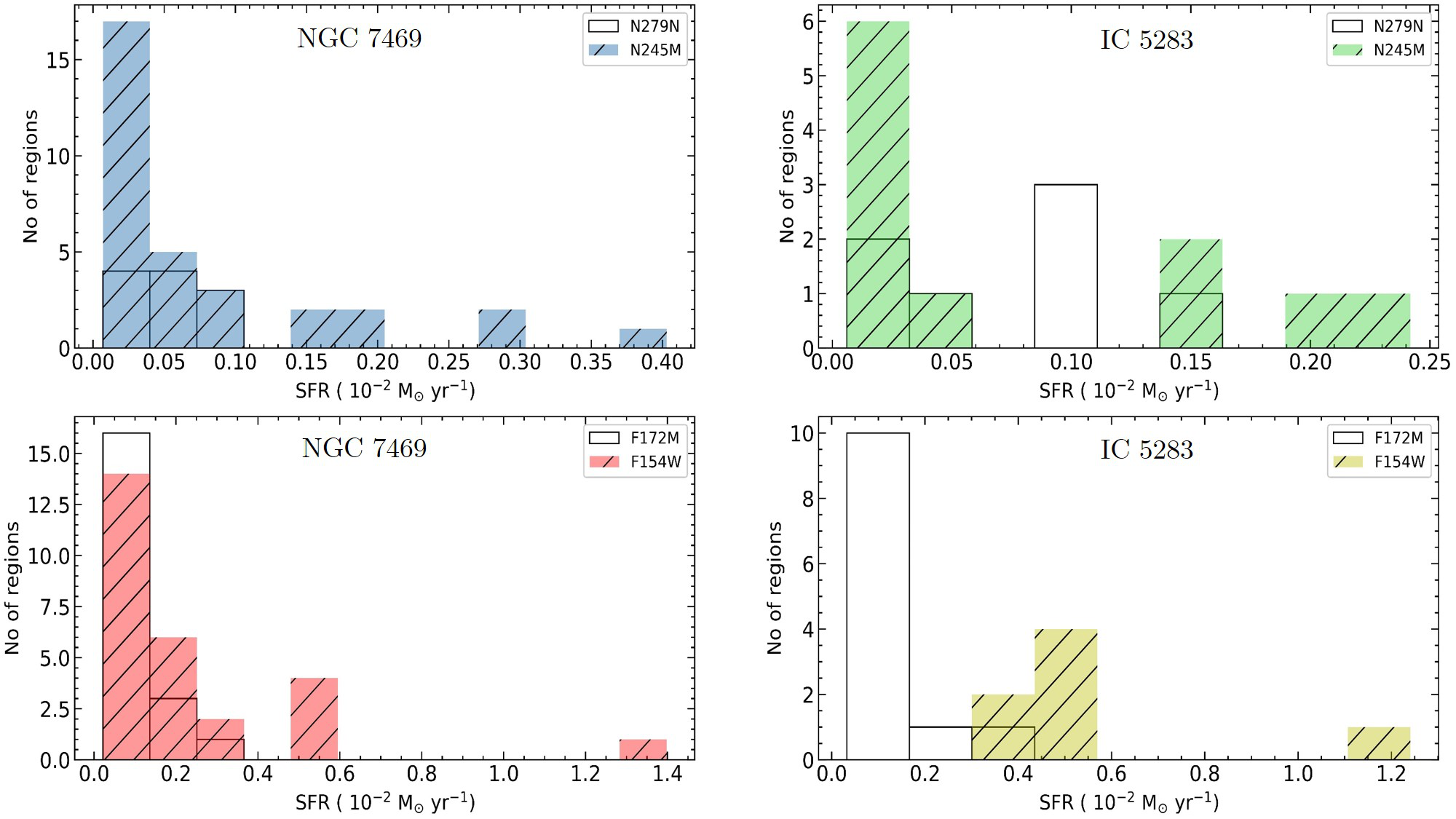}
    \caption{Distribution of SFR of NGC~7469 and IC~5283 in various UVIT filters}
    \label{Merged_SFR}
\end{figure*}
\subsubsection{Sizes of the Star-Forming regions}

SEXTRACTOR gives us the lengths of semimajor and semiminor axes in pixel coordinates and position angles in degrees corresponding to each elliptical-shaped SF region. We converted these to arcseconds and then to kpc using the redshift value of $z=0.0163$. We noticed that the areas of the SF regions detected in UV (N245M, F154W, N279N, and F172M) vary from $0.17 \text{ kpc}^2$ to $3.24 \text{ kpc}^2$ in the FUV bands and $0.24 \text{ kpc}^2$ to $5.68 \text{ kpc}^2$ in the NUV bands. The distribution of the areas of the SF regions (segregated by filters and galaxies) is shown in Figure \ref{Merged_Knots}. The median value of the SF regions area are $ 1.24 \text{ kpc}^2$ for the N245M filter, $ 1.31 \text{ kpc}^2$, $ 1.61 \text{ kpc}^2$ for the F154W filter, and $ 0.71 \text{ kpc}^2$ for the F172M filter. Other studies have also identified and calculated the sizes of the SF regions in different galaxies using UVIT. For example, for a nearer galaxy NGC~2336 ($z= 0.00735$) studied by \citet{2018MNRAS.481.1212R}, the median distribution of the SF sizes comes out to be $ 0.63 \text{ kpc}^2$ and  $ 0.48 \text{ kpc}^2$ respectively in F154W and N242W filters (assuming circular SF regions) based on the results given in \citet{2018MNRAS.481.1212R}. 
The UVIT, however, does much better for the dwarf Seyfert galaxy NGC~4395 which is much closer at z = 0.00106 and has been studied by \citet{2023ApJ...950...81N}. They have reported the SF regions sizes using the the F148W filter to be varying from $ 8.6 \times 10^{-4} \text{ kpc}^2$ to $ 488.1 \times 10^{-4} \text{ kpc}^2$ with a median of $ 40.2 \times 10^{-4} \text{ kpc}^2$. 

In some of the filters, there are regions where multiple clumps are identified as one larger clump in IC~5283. This might have several implications for the interpretation of observational data and the analysis of star formation activity in galaxies. We tried to reduce this as much as possible, as the presence of such structures could lead to overestimation or underestimation of the fluxes/apparent magnitudes of such regions.

The distribution of the sizes of the SF regions is different in both galaxies. We also see a tidal tail in IC~5283. Although a similar SFR distribution, \citet{1994AJ....108...90M} suggests that the galaxies have produced very different responses to the interaction, in particular regarding the distribution of the star formation regions which is true in our case too.

\subsubsection{Other Parameters of Star-Forming Regions}
\label{sec::sfr}

The fluxes for the star-forming regions were derived from the observed counts (derived from SEXTRACTOR) using the conversion factors given in the \citep{2020AJ....159..158T} and \url{https://uvit.iiap.res.in/Instrument/Filters} implemented in Python code. The positions, calculated flux values, and the distance {\footnote{the distance of the SF regions from the central AGN is only calculated in the case of NGC~7469 as the identification of the center in the case of IC~5283 is rather complex.}} of the SF regions from the central AGN in each UVIT filter for NGC~7469 and IC~5283 are presented in the electronic format in the CDS.

\begin{figure*}
\centering
	\includegraphics[width=1.0\textwidth]{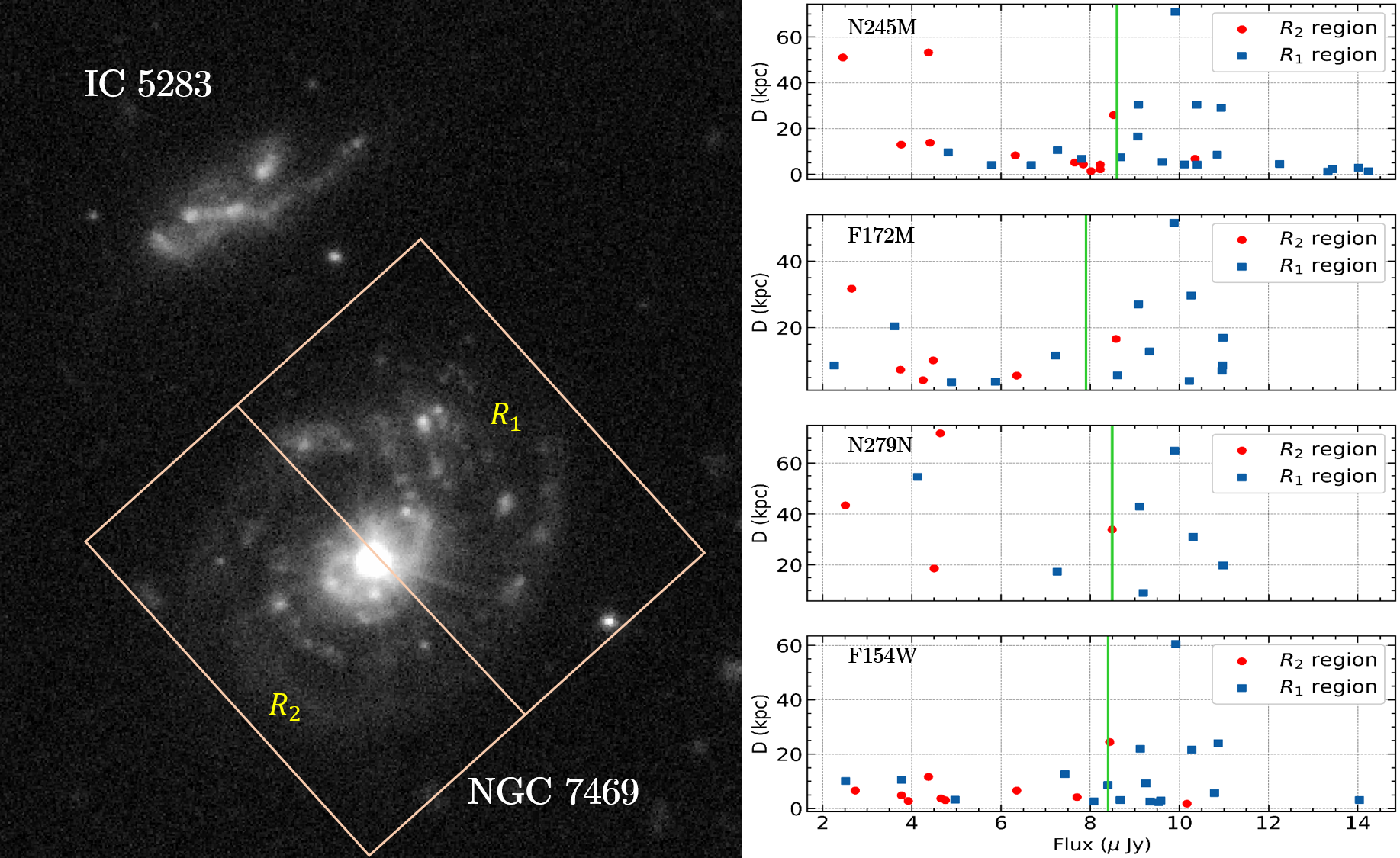}
    \caption{(left) Star-forming regions (figure \ref{Merged_photometry}) of the galaxy NGC~7469 (N245M filter) divided by the areas - $R_1$ and $R_2$ containing Northern arms and Southern arms respectively. The measurements of the box size and division are provided in Section \ref{sec::sfr}. North is up and West is left direction. (right) Plot showing the distance from the central AGN as a function of flux values (in $\mu$Jy) of the star-forming regions (refer the electronically available Tables) in the $R_1$ and $R_2$ areas shown by blue squares and red circles respectively in various UVIT filters. The green line is the median flux value of the SF regions in those filters. It clearly shows the enhanced star-formation process occurring in the $R_1$ region suggesting that the interaction might have occurred along that arm.}
    \label{Merged_flux_reg}
\end{figure*}

The SF regions exhibit varying sizes and levels of UV emission, indicative of ongoing star formation activity. 
The Far-Ultraviolet (FUV) light, spanning the wavelength range from 120 to 320 nm, is predominantly emitted by stars with masses of several solar masses and mean ages ranging from 10 to 100 million years (Myr) \citep{2011ApJ...741..124H, 2011ApJ...737...67M, 2012ARA&A..50..531K}. This emission is closely associated with recent star formation activities within a galaxy \citep{2006ApJ...648..987P}.  

We have estimated the star formation rate (SFR) in the SF regions by using the prescription provided by \citet{2006ApJS..164...38I} and \citet{2016MNRAS.461.1898W} based on the FUV and NUV luminosities. The distribution of the SFR in the SF regions is provided in Figure \ref{Merged_SFR}.
This approach takes into account the known relationship between the FUV and NUV emission and the presence of young, massive stars and consequently derives an estimate of the recent star formation rate within the galaxy under investigation.

The FUV and NUV star formation rates are given by
\begin{equation*}
\centering
    \text{log}(\text{SFR}_{\text{FUV}}/\text{M}_{\odot} \text{yr}^{-1}) = \text{log}(\text{L}_{\text{FUV}}/\text{L}_{\odot}) - 9.51
\end{equation*}

\begin{equation*}
\centering
    \text{log}(\text{SFR}_{\text{NUV}}/\text{M}_{\odot} \text{yr}^{-1}) = \text{log}(\text{L}_{\text{NUV}}/\text{L}_{\odot}) - 9.58
\end{equation*}

where $\text{L}_{\text{FUV}}$ and $\text{L}_{\text{NUV}}$ are the luminosities in the F154W and F172M filters and the N245M and N279N luminosities respectively of the galaxy corrected for Galactic extinction ($A_{\text{FUV}}$ ) and ($A_{\text{NUV}}$ ). 
The calibration for this prescription is derived from Starburst99 models \citep{1999ApJS..123....3L}, assuming a solar metallicity and a Salpeter Initial Mass Function (IMF) from
$0.1-100M_{\odot}$. Galactic extinction, $A_{V} = 3.1 \times E(B-V)$  \citep{2018A&A...615A..72M}, where the colour excess $E(B-V ) = 0.061$ for {NGC~7469 and IC~5283} is obtained from the dust reddening maps of \citet{2011ApJ...737..103S}.

We also investigate if the interaction between the two galaxies might have occurred along one of the directions. We created a box of length and width of $1.5'$ and $1.55'$ respectively at an angle of $130^{\circ}$ centered on the central AGN source (RA, Dec) = (23:03:15.49, +8:52:29.03). We draw a line parallel to the width passing through the central AGN with coordinates (RA, Dec) = (23:03:15.60, +8:42:26.00). This way we split the NGC~7469 into two halves - $R_1$ and $R_2$ containing the Northern arms and Southern arms respectively, and plotted the flux values (in $\mu$Jy) in various UVIT filters as a function of distance from the central AGN for both the divided regions. It is shown in the Figure \ref{Merged_flux_reg}. The flux values for the SF regions in the regions $R_1$  and $R_2$ are shown by blue squares and red circles respectively. The green line shows the median of the overall flux values. We have tabulated the flux values in the Tables available electronically at the CDS portal. One can see that the star-forming regions in the $R_1$ area (refer to Figure \ref{Merged_flux_reg}) have higher flux values regions (and hence SFR) compared to the $R_2$ region, in all of the UVIT filters. There is a clear indication that there is an enhanced star formation process going on in the outer arms lying towards the northwest side of the NGC~7469 which supports the theory that the interaction between NGC~7469 and the IC~5283 might have triggered the star-formation process, similar to a conclusion made by \citet{2002ApJS..143...47D}. It is possible that the interaction might have taken place from the direction of the northwest side of NGC~7469 as it was in close vicinity of the IC~5283 galaxy.

We note that obscuration by dust internal to the galaxy can affect observations in the ultraviolet (UV) wavelengths and may have several consequences on the derived properties of galaxies, such as the estimated number of clumps, their sizes, and star formation rates (SFR) \citep{2001PASP..113.1449C, 2011MNRAS.410.2291W}. Since UV emission is often associated with regions of active star formation, such as young stellar clusters or star-forming regions, it can obscure these regions, making them less visible or undetectable in UVIT data \citep{2023MNRAS.520.3712S}. This can lead to an underestimate of the number of detected star-forming clumps or an incomplete census of star-formation activity within the galaxy, the underestimation of UV flux, and consequently the SFR values.

\begin{table*}
\centering
	\caption{Results obtained from Kolmogorov-Smirnov test for the comparison of SFR distributions in NGC~7469 and IC~5283 in different filters}
	{
	\begin{tabular}{lcc}
		\hline
	Filter & KS Statistic & p-value \\
		\hline	
            \hline
        N245M &  0.267 & 0.498\\     
	F154W &  0.814 & 0.001\\
        N279N &  0.488 & 0.181\\
	F172M &  0.216 & 0.806\\ 
  	
\hline  
	\end{tabular}}
 \label{KS_table}
\end{table*}

\subsection{Kolmogorov-Smirnov (KS) Tests}
\label{KS_test}

We employed the Kolmogorov-Smirnov (KS) test \citep{Shiryayev1992} to assess the similarity or dissimilarity between the SFR's distributions in the NGC~7469 and IC~5283. The KS test was performed separately for all the UVIT filters and the results are presented in Table \ref{KS_table}. 
The lowest p-value is seen in the F154W filter which has the widest wavelength band of $140-175$ nm. The F154W filter has the highest sensitivity for the Helium recombination line (HeII) at $164$ nm, OIII line at $166.3$ nm and NII line at $173.7$ nm which are strong indicators of massive star formation (see also \cite{2020Galax...8...13L}), especially the He recombination line \cite{2021MNRAS.503.6112S}. This suggests that the distribution of massive O/WR  stars in these galaxies is different. On the other hand, the other FUV filter F172M is comparatively narrower and with a lower efficiency. The NUV filters are much narrower in their wavelength range and cover only weaker lines of NeIV, MgII considered as the indicators of star formation activity. From the KS test results, there is no significant difference in the SFR distributions between the two galaxies' star-forming regions in the N245M, F172M, and N279N filters. The similar distribution of SFR in three filters indicates that, overall, the star formation processes in both NGC 7469 and IC 5283 are comparable. This might be attributed to factors such as the overall gas content, density, or gravitational interactions between the galaxies, which influence the star formation rates consistently across multiple filters. However, understanding the physics driving the low p-value in the F154W filter requires further analysis and possibly additional observational data. 


\begin{table*}
	\centering
	\caption{Integrated flux measurements for NGC~7469 and IC~5283 in different wavebands, in increasing order of wavelength from top to bottom.}
	\resizebox{2\columnwidth}{!}{
	\begin{tabular}{lccccc} 
 	\hline
	 &   & NGC~7469 &  & IC~5283 & \\
	\hline \hline
	Filter & Effective $\lambda$ (Å) & Flux with Uncertainty (mJy) & Source & Flux with Uncertainty (mJy) & Source\\
	\hline	
	\textit{AstroSat} F154W & 1.54E+3 &  3.160$\pm$0.007 & measured from image  & 0.160$\pm$0.001 & measured from image \\

	\textit{AstroSat} F172M & 1.72E+3 &3.91$\pm$0.01 & measured from image  &0.240$\pm$0.006 & measured from image \\
 
	\textit{AstroSat} N245M & 2.45E+3 & 6.480$\pm$0.003 & measured from image &0.380$\pm$0.001 & measured from image\\	

    \textit{AstroSat} N279N  & 2.79E+3  & 8.72$\pm$0.02 & measured from image   &0.630$\pm$0.006 & measured from image\\

    SDSS \textit{u} & 3.54E+3 & 13.3$\pm$0.2 & measured from image  &1.00$\pm$0.07 & measured from image\\
    
 SDSS \textit{g} & 4.77E+3  & 31.5$\pm$0.3 & measured from image  &5.58$\pm$0.08 & measured from image\\

   SDSS \textit{r} & 6.23E+3 & 44.3$\pm$0.4 & measured from image  &6.30$\pm$0.10 & measured from image\\
   
  SDSS \textit{i} & 7.62E+3 & 76.4$\pm$0.5 & measured from image  &8.91$\pm$0.14 & measured from image\\

SDSS \textit{z} & 9.13E+3  & 98.3$\pm$0.6 & measured from image  &13.71$\pm$0.21 & measured from image\\

2MASS & 1.23E+4 & 69.7$\pm$0.5 & measured from image  &8.51$\pm$0.12 & measured from image\\

WISE W1 & 3.35E+4 & - & - & 13.4 $\pm$ 0.3& ViZer Catalog \\

WISE W2 & 4.60E+4 & - & - & 8.9 $\pm$ 0.2& ViZer Catalog\\

WISE W3 & 1.16E+5 & - & - & 69.7 $\pm$ 0.9& ViZer Catalog \\
    
IRAS $12 \mu$ m & 1.20E+5&  1590$\pm$39 & \citet{2003AJ....126.1607S} & - & -\\
    
IRAS $25 \mu$ m & 2.50E+5 & 5960$\pm$32 & \citet{2003AJ....126.1607S}  & -&-\\
    
IRAS $60 \mu $  m& 6.00E+5 & 2730$\pm$40 & \citet{2003AJ....126.1607S} & - & - \\
    
IRAS $100 \mu $ m & 1.00E+6 & 35100$\pm$599 & \citet{2003AJ....126.1607S}  & - & - \\

SCUBA 850 $\mu$ m & 8.55E+6 & 264$\pm$30 & \citet{scuba}  & - & - \\

   \hline
	\end{tabular}}
	\label{SEDTable}

    	\tablefoot{ The central numbers of the UVIT filters i.e. 245, 279, 154, and 172 correspond to the central wavelength of the corresponding filter in nm.}
\end{table*}

\begin{figure*}
\centering
	\includegraphics[width=1.0\textwidth]{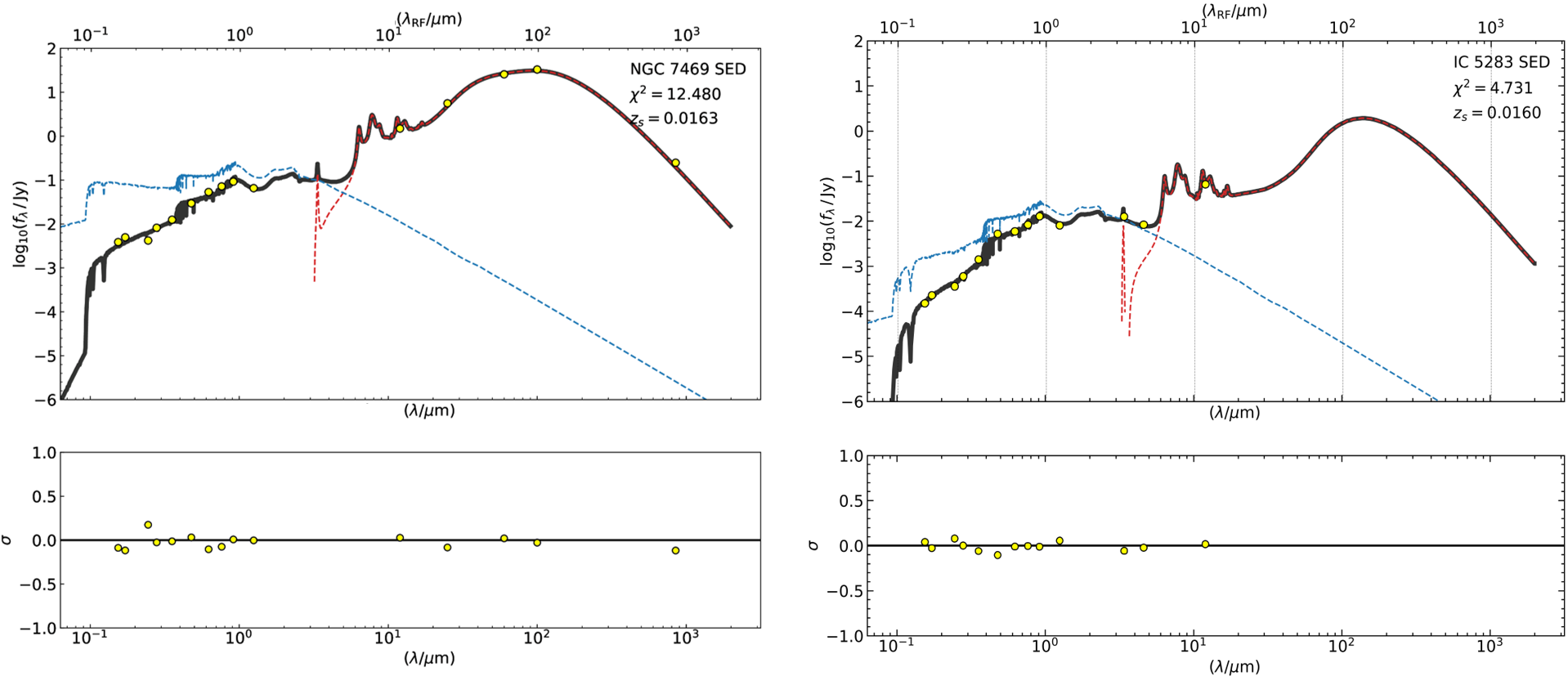}
    \caption{{The integrated UV to near-IR spectral energy distribution
(SED) of the galaxies NGC~7469 and IC~5283. In this two-panel plot, the top panel shows the photometric data from Table \ref{SEDTable} for the two galaxies are shown as yellow dots which were used to fit a SED using the code MAGPHYS \citep{2011ascl.soft06010D}.   The blue dashed line represents the unabsorbed stellar component, while the red dashed lines show the dust re-emission.
The black solid line represents the total emission. The bottom panel shows the residuals concerning the best fit. The parameters $\chi^2$ and $z_s$ represent the chi-square of the best-fit model and the redshift of the galaxy respectively. Table \ref{SED_fit_Table} provides the extracted physical parameters derived from this SED fitting.}}
    \label{SED}
\end{figure*}

\begin{table*}
\centering
	\caption{A comparison of NGC~7469 and IC~5283 with respect to the physical parameters estimated with the SED fitted using the MAGPHYS code \citep{2008MNRAS.388.1595D}.}
	{
	\begin{tabular}{lcc} 
 \hline
 & NGC~7469 & IC~5283\\
		\hline \hline
	Main Parameters & Values & Values \\
		\hline	
	log \ SFR ($\text{M}_{\odot} \ yr^{-1}$)  & $1.53 \pm 0.08$ & $0.57 \pm 0.14$\\

	log\ $\text{M}_{*}$  ($\text{M}_{\odot}$) & $10.80 \pm 0.17$ & $10.31 \pm 0.09$ \\
     
     log \ $\text{M}_\text{Dust}$ ($M_{\odot}$) & $7.71 \pm 0.12$ & $7.13 \pm 0.20$ \\
        
        log \ sSFR ($yr^{-1}$) &  $-9.27 \pm 0.20$ & $-9.74 \pm 0.24$ \\
		\hline  
	\end{tabular}}
	\label{SED_fit_Table}
\end{table*}

\section{Spectral Energy Distributions}

We obtained the SED of the integrated emission of the galaxies by combining the UV data with the optical and infrared data from the archives as detailed below.

\subsection{Optical Data}
\label{opti}
The optical photometric data utilized in this study were obtained from the Sloan Digital Sky Survey (SDSS), specifically from data release 14 \citep{1995ApJ...438..604M}. The SDSS images were exposed for a duration of 53.9 seconds in each of the five filters: u, g, r, i, and z. 
Photometric measurements were carried out in all four wavebands, and the resulting counts were subsequently converted to fluxes in Jansky (Jy) using the standard conversion factor.

\subsection{Infrared Data}
\label{infra_red}
The infrared photometric data for NGC~7469 and IC~5283 were obtained from the various surveys and are presented in Table \ref{SEDTable}. The near-infrared data included images from WISE W1, W2, and W3. Moreover, 2MASS images used in this study were acquired through the J filter. Far-infrared data have been acquired from surveys like IRAS and SCUBA \citet{2003AJ....126.1607S} and \citet{scuba}. Our flux measurements (provided in Table \ref{SEDTable}) match pretty well with the existing literature values \citep{2003AJ....125..525J}.

The integrated flux measurements of the entire galaxy, NGC~7469, in different UV filters, were obtained by using a photometric radius of $0.9'$ to study its SED. Similarly, $0.55'$ and $0.3'$ were the values chosen as semi-major and semi-minor axis respectively for flux measurements and for SED analysis of the IC~5283 galaxy. The measured integrated flux values for the two galaxies are given in Table \ref{SEDTable}. The SEDs of the integrated galaxy emission for NGC~7469 and IC~5283 are shown in Figure \ref{SED}. 

\subsection{Modelling of the SED}
SED was analysed using the MAGPHYS (Multi-wavelength Analysis of Galaxy Physical Properties) code \citep{2008MNRAS.388.1595D}. 
The MAGPHYS code utilises a combination of stellar population synthesis models and dust emission models to constrain various physical parameters of galaxies. It combines BC03 optical/NIR stellar models, which includes the effects of dust attenuation as described in \cite{2000ApJ...539..718C}, and the MIR/FIR dust emission computed as in \citep{2008MNRAS.388.1595D}, linking the two components through energy balance: the total energy absorbed by dust in stellar birth clouds and in the ambient interstellar medium is re-distributed at infrared wavelengths.

The SED fit uses the observed flux densities from ultraviolet (UV) to far-infrared (FIR) wavelengths (see Table \ref{SEDTable}) with a grid of model SEDs that span a range of star formation histories, stellar masses, dust properties, and other parameters.

By comparing the observed flux densities with the model grid, we derived the best-fit values for several key physical parameters like the star formation rate (SFR), stellar mass ($\text{M}_{*}$), dust mass ($\text{M}_\text{Dust}$), and specific star formation rate (sSFR) and the procedure is outlined below. These values are listed in Table 
\ref{SED_fit_Table}. The best-fit models for the SEDs of the integrated galaxy emission for NGC~7469 and IC~5283 are shown in Figure \ref{SED}. 

Star formation history is an important tool for understanding the intrinsic evolution of a galaxy and the effects of interactions with its neighbouring systems \citep{2014MNRAS.442.1897R}. We have extracted the total star-formation rate (SFR) of the two galaxies using MAGPHYS \citep{2012ARA&A..50..531K} and by combining the UV and infrared emissions, which are sensitive to young and dust-enshrouded star formation respectively. The MAGPHYS code utilises stellar population models to estimate the stellar mass by matching the observed SED with models that account for various stellar ages and metallicities \citep{2011ApSS.331....1W}. On the other hand, the dust absorbs the ultraviolet (UV) emission of young stars, allowing gas to cool and condense to form new stars. 
 
Dust emission (infrared (IR) and sub-millimetre (sub-mm) regimes) is often used to trace the ongoing rate of star formation \citep{2010A&A...518L..24N, 2012ApJ...745..182N, 2011A&A...533A.119E, 2014MNRAS.444.1647K}. 

We determined the dust mass ($M_{\text{Dust}}$) in NGC~7469 and IC~5283 by incorporating dust emission models in MAGPHYS.
This is the amount of dust available for obscuring starlight and potentially fuelling star formation processes \citep{2010A&A...523A..78D}.
We also derived the specific star formation rate (sSFR) for NGC~7469. The sSFR represents the star formation rate normalised by the stellar mass and provides a measure of the relative level of ongoing star formation activity compared to the existing stellar population. It is a key quantity in characterising the evolutionary state of galaxies, with high sSFR values indicating more intense star formation relative to the stellar mass \citep{2011A&A...533A.119E}.

The ratio of stellar masses of the two galaxies is $0.3$ (IC~5283/NGC~7469) suggesting that these galaxies are undergoing a major merger. In \cite{2016ApJ...825..128L}  merger sequence is listed based on the visual classification of different interacting galaxy systems, including NGC~7469 and IC~5283. In this sequence, the merger is denoted by M1 to M5 with an increasing number corresponding to later stages of mergers. Another criterion of the merger status is the projected distance,  a distance larger than $10$ kpc corresponds to interacting systems at stages M1 (galaxies in initial approach) and M2 (visible tidal tails).
The visual inspection classifies this galaxy system as major merger stage 2 (M2) where the interacting galaxy pairs have prominent tidal bridges and tails and are consistent with having already undergone a pericentre passage (see also \cite{2020MNRAS.492.2075B}).
We find the projected nuclear separation to be $26.59$ kpc which is consistent with the value of 27.2$\pm$4.1 kpc quoted for the M2 stage by \citet{2016ApJ...825..128L} and with the separation of $26.2$ kpc estimated by \citet{2021ApJS..257...61Y} reiterating that the galaxies are indeed undergoing a major merger. We compute specific star formation rate of $0.5~ Gyr^{-1}$ for NGC~7469 and $0.1~Gyr^{-1}$ for the companion galaxy. 

Our SED fitting results show SFR of $\sim 35 / M_{\odot}/ yr^{-1}$ for the entire galaxy, whereas our photometric measurements show aggregated SFR of $\sim 1-2 \ M_{\odot}/ yr^{-1}$ since we masked the central SB, CND, and the AGN, and only considered the outer regions. This result is consistent in that most of the star formation in NGC 7469 is confined to a clumpy starburst ring, but the star formation efficiency remains quite elevated even for the nuclear region that is most affected by the AGN \citet{2023ApJ...953L...9Z}.

Interactions enhance Specific star formation rates  \citep{2020MNRAS.494.4969P}  and the enhancements extend to projected separations of $260$ kpc (for physical distances it is $280$ kpc)  as seen in  IllustrisTNG simulations \citep{2019ComAC...6....2N} and Sloan Digital Sky Survey (SDSS) galaxies. The projected distance, despite being a useful indicator does not give information on the actual distance between centres of the galaxies. Detailed dynamical simulation of the galaxy system will reveal more details about this particular system.

\section{Summary}
\label{discussions}

In this paper, we present detailed images of the galaxy pair NGC~7469 and IC~5283 in four wavelength bands of FUV and NUV, focusing mainly on the pair as a whole and the interaction among the galaxies.  
The galactic arms NGC~7469, in particular, the inner arm (also known as the outer ring)  and the outer spiral arm are clearly visible in the composite image with a resolution that is better than the previous UV observations by GALEX and XMM-OM.
We have detected star-formation knots in the spiral arms of the galaxy NGC~7469 and also detected a few star-forming knots in the companion IC~5283.

We have presented the results of photometry of the star-forming regions in the spiral arms of the NGC~7469 and the star-forming knots in the companion IC~5283 based on the UVIT observations in various filters. These were analysed with the intention of studying the interaction between these galaxies using star-forming regions. We were able to identify a total of 51 SF regions, out of which 33 were new regions that could comprise star clusters, dense HII regions, etc. Other 18 were previously observed by various observing telescopes including HST, VLA, and Gemini.
We see there is an enhanced star formation process in the outer arms which are towards the northwest side of the NGC~7469. Many star-forming knots are also present in IC~5283. This is consistent with the presence of interaction between these galaxies and this interaction may have triggered star formation in this arm. 

To compare the star formation distributions in the two galaxies, we performed Kolmogorov-Smirnov tests. 
The results show very little differences of significance in the SFR distributions of NGC~7469 and IC~5283. There is some indication, however, that the interaction between the galaxies may have influenced their star formation processes differently.
The surface brightness of NGC~7469  in the far-UV filters (F154W and F172M) and in one of the near-UV filters (N279N) is higher in the case of NGC~7469 compared to IC~5283 whereas surface brightness value in the N245M filter is comparable for both galaxies.

The stellar mass ratio of the galaxies confirms that they are undergoing a major merger. A comparison with simulations indicates that the current stage of the merger is post-first pericentre transit and simulations of the system will reveal the exact nature of the dynamics. It is to be noted that most of the SFR of the galaxy NGC 7469 is confined within the central SB ring, CND, and near the central AGN. However, the enhanced star formation in the outer regions is likely due to the merger. The results are consistent with previous observations and support the argument that these galaxies are indeed, interacting. 

The SFR obtained for our interacting pair appears to be higher than that obtained in studies of other starburst galaxies. For example, a study by \citep{2023RNAAS...7..232W} for $~30$ galaxies found that the average SFR for the star-forming galaxies is around $8-9 \ M_{\odot} Yr^{-1}$. Even galaxies in an interacting pair, NGC~4676a/NGC~4676b, show an SFR of $6.2 \ M_{\odot} Yr^{-1}$ for NGC~4676a and $2.3 \ M_{\odot} Yr^{-1}$ for NGC~4676b based on FUV and mid-IR data. However, some interacting/colliding galaxies with ages in the range $\sim 100 \ MYr$, like the galaxy pair, NGC~2207/IC~2163 do show a higher SFR of $23.7 \ M_{\odot} Yr^{-1}$ using X-ray data.  Similarly galaxies in another distant interacting pair, like Arp~299 (NGC~3690/IC~694) show SFR of $75 \ M_{\odot} Yr^{-1}$ using X-ray data.

Further investigations, such as spatial analysis and multi-wavelength studies, are warranted to better understand the specific mechanisms driving the similarity in star formation and to explore the implications for the overall evolution of these galaxies.

\section*{Acknowledgements}
We thank the Indian Space Research Organisation for scheduling the observations within a short period of time and the Indian Space Science Data Centre (ISSDC) for making the data available.  Kulinder Pal Singh
thanks the Indian National Science Academy for support under the INSA Senior Scientist Programme. This work has been performed utilising the calibration databases and auxiliary analysis tools developed, maintained, and distributed by {\it AstroSat}-SXT team with members from various institutions in India and abroad and the  SXT Payload Operation Center (POC) at the TIFR, Mumbai for the pipeline reduction. The work has also made use of software, and/or web tools obtained from NASA's High Energy Astrophysics Science Archive Research Center (HEASARC), a service of the Goddard Space Flight Center and the Smithsonian Astrophysical Observatory. Abhinna Sundar Samantaray also thanks the Hector Fellow Academy (HFA) for the support with the research grants for a part of this project.

\section*{Facilities:} 
\textit{AstroSat}

\section*{ORCID IDs}
Abhinna Sundar Samantaray \orcidlink{0000-0002-0635-2264}\\
H. K. Jassal \orcidlink{0000-0003-2486-5634}\\
Kulinder Pal Singh \orcidlink{0000-0001-6952-3887}\\
Gulab Chand Dewangan \orcidlink{0000-0003-1589-2075}

\section*{Data Availability:}
{\it AstroSat} data for all the observations used in this paper are publicly available from the {\it AstroSat} archives maintained by the ISSDC, Bengaluru: \url{https://astrobrowse.issdc.gov.in/astro_archive/archive/Home.jsp}.

%
%
\bibliographystyle{aa}
\bibliography{aanda}

\end{document}